\documentclass[pra,aps,superscriptaddress,showpacs]{revtex4}
\usepackage[dvips]{graphicx}
\begin{document}
\author{Jing-Ling Chen}
\email{chenjl@nankai.edu.cn} \affiliation{Theoretical Physics
Division, Chern Institute of Mathematics, Nankai University, Tianjin
300071, People's Republic of China} \affiliation{Department of
Physics, National University of Singapore, 2 Science Drive 3,
Singapore 117542}
\author{Chunfeng Wu}
\affiliation{Department of Physics, National University of
Singapore, 2 Science Drive 3, Singapore 117542} \affiliation{School
of Physics, Northeast Normal University, Changchun 130024, People's
Republic of China}
\author{L. C. Kwek}
\affiliation{Department of Physics, National University of
Singapore, 2 Science Drive 3, Singapore 117542} \affiliation{Nanyang
Technological University, National Institute of Education, 1,
Nanyang Walk, Singapore 637616}
\author{C. H. Oh}
\email{phyohch@nus.edu.sg} \affiliation{Department of Physics,
National University of Singapore, 2 Science Drive 3, Singapore
117542}
\title{Bell inequalities for three particles}
\begin{abstract}
We present new tight Bell inequalities expressed by probabilities
for three four- and five-dimensional systems. The tight structure of
Bell inequalities for three $d$-dimensional systems (qudits) is
proposed. Some interesting Bell inequalities of three qubits reduced
from those of three qudits are also studied.
\end{abstract}
 \maketitle

\section{Introduction}
That no local and realistic theory agrees with all predictions of
quantum mechanics was shown by Bell in 1964 \cite{bell} through the
violations of certain constraints. Local realism imposes constraints
in the form of Bell inequalities on statistical correlations of
measurements on multiparticles. Quantum mechanics predicts
violations of such Bell inequalities. The original Bell inequality
and the subsequent famous CHSH inequality \cite{chsh}, the latter
being cast into a form more amenable for experimental verification,
were formulated for the simplest composite quantum system, namely, a
system of two qubits.

Bell inequalities which eliminate local realistic description are of
importance not only on fundamental research but on identifying
ultimate resources for quantum information processing. It was shown
that there is a direct link between the security of quantum
cryptography and the violation of Bell inequalities \cite{Artur,
Acin2}. Collins \textit{et al} \cite{Gisin} found a tight Bell
inequality for two arbitrary $d$-dimensional systems (or two qudits)
in terms of joint probabilities, hereafter we call it as the CGLMP
inequality. For $d=2$, the CGLMP inequality reduces to the CHSH
inequality. Alternatively, for $N$ particles of two dimensions
(called $N$-qubit), it was shown that there exists a general Bell
inequality which is a sufficient and necessary condition for
$N$-body correlations to be describable in a local and realistic
theory based on two local settings for each observer \cite{ZB}.
However, Bell inequalities for $N$ ($N>2$) entangled $d$-dimensional
($d>2$) quantum systems are not so well formulated as those for
two-qudit or $N$-qubit. Only recently, the problems have been solved
partly in the case of three three-dimensional particles in
\cite{Acin}. The authors developed a coincidence Bell inequality in
terms of probabilities. For general $N$ ($N>2$) entangled
$d$-dimensional quantum systems, no Bell inequality for either
probabilities or correlation functions has been presented until now,
although GHZ paradox has been generalized to $N$ qudits systems
\cite{paradox}.

Even for $N$-qubit systems, there is one problem on Bell inequality
remains, that is ``do all pure entangled states violate Bell
inequalities for correlation functions"? In other words, it is that
whether the theorem of Gisin \cite{gisin,popescu} can be generalized
to $N$ qubits or not. It is found that there is a family of pure
entangled states of $N$ qubits which do not violate all Bell
inequalities \cite{Zukowski}. For three qubits, we have proposed a
Bell inequality to solve the problem \cite{3qubit}.

In this work, we present the tight Bell inequalities expressed by
probabilities for three four- and five-dimensional systems. The
tight structure of Bell inequalities for three $d$-dimensional
systems (qudits) is proposed. Some interesting Bell inequalities for
three qubits reduced from those of three qudits are also studied.

\section{Bell inequalities for three $d$-level systems}

Local realism cannot exhibit arbitrary correlations. The constraints
that local realistic correlations must obey can be written in the
form of Bell inequalities. For a three $d$-dimensional system with
an arbitrary value of $d$, some efforts have been given to develop
Bell inequalities recently. The first step came in 1990 with a paper
of Mermin \cite{Mermin} in which he derived a Bell inequality for
arbitrary $N$-qubit states; quantum mechanics violates this
inequality by an amount that grows with $N$. This result clearly
gives us a first three-qubit Bell inequality in a correlation form,
\begin{eqnarray}
Q_{112}+Q_{121}+Q_{211}-Q_{222}\leq 2.
\end{eqnarray}
It can be expressed in terms of probabilities,
\begin{eqnarray}\label{3qubitinequality}
&&P(a_1+b_1+c_2=0)-P(a_1+b_1+c_2=1)+P(a_1+b_2+c_1=0)-P(a_1+b_2+c_1=1) \nonumber \\
&&+P(a_2+b_1+c_1=0)-P(a_2+b_1+c_1=1)-P(a_2+b_2+c_2=0)+P(a_2+b_2+c_2=1) \leq 2.
\end{eqnarray}
This Bell inequality is maximally violated by the three-qubit GHZ
state $|\psi_2\rangle=\frac{1}{\sqrt{2}}(|000\rangle+|111\rangle)$.
But for the generalized GHZ states
$|\psi_2\rangle_{GHZ}=\cos\xi|000\rangle+\sin\xi|111\rangle$, there
exists one region $\xi \in(0,\pi/12]$ in which the Bell inequality
is not violated.

The second step is due to Ref. \cite{Acin}. The authors developed a
three-qutrit Bell inequality involving probabilities which can be
given in an alternative form,
\begin{eqnarray}\label{3qutriitinequality}
&&-P(a_1+b_1+c_1=0)-P(a_1+b_1+c_1=1)+2P(a_1+b_1+c_1=2)+P(a_1+b_1+c_2=0)  \nonumber \\
&&-2P(a_1+b_1+c_2=1)+P(a_1+b_1+c_2=2)+P(a_1+b_2+c_1=0)-2P(a_1+b_2+c_1=1) \nonumber \\
&&+P(a_1+b_2+c_1=2)+P(a_2+b_1+c_1=0)-2P(a_2+b_1+c_1=1)+P(a_2+b_1+c_1=2)  \nonumber \\
&&+2P(a_1+b_2+c_2=0)-P(a_1+b_2+c_2=1)-P(a_1+b_2+c_2=2)+2P(a_2+b_1+c_2=0) \nonumber \\
&&-P(a_2+b_1+c_2=1)-P(a_2+b_1+c_2=2)+2P(a_2+b_2+c_1=0)-P(a_2+b_2+c_1=1) \nonumber \\
&&-P(a_2+b_2+c_1=2)-2P(a_2+b_2+c_2=0)-2P(a_2+b_2+c_2=1)+4P(a_2+b_2+c_2=2) \leq 6.
\label{3qutritinequality}
\end{eqnarray}
The above inequality is maximally violated by the three-qutrit GHZ
state
$|\psi_3\rangle=\frac{1}{\sqrt{3}}(|000\rangle+|111\rangle+|222\rangle)$.
It is worthy of mentioning that both inequalities
(\ref{3qubitinequality}) and (\ref{3qutritinequality}) are tight.

Here the third step is coming. Our approach of constructing a new
Bell inequality for tripartite four-dimensional systems is based on
the Gedanken experiment. There are three separated observers,
denoted by A, B, and C hereafter, each can carry out two possible
local measurements, $A_1$ or $A_2$ for A, $B_1$ or $B_2$ for B and
$C_1$ or $C_2$ for C respectively. Each measurement may have four
possible outcomes, labeled by 0, 1, 2 and 3. We denote the
observable $X_i$ measured by party $X$ and the outcome $x_i$ with
$X=A, B, C\; (x=a, b, c)$. A local realistic theory can be described
by $8\times 56$ probabilities. Here we denote the joint probability
$P(a_i+b_j+c_k=r)$ that the measurements $A_i$, $B_j$ and $C_k$ have
outcomes that differ, modulo four, by $r$:
\begin{eqnarray}
P(a_i+b_j+c_k=r)=\sum_{a,b=0,1,2,3}P(a_i=a,b_j=b,c_k=r-a-b).
\end{eqnarray}
Some of the local realistic constraints are trivial, such as
normalization and the no-signaling conditions which are not violated
by quantum predictions. Only the non-trivial inequality, which is
not true for quantum mechanics, is of use for checking whether we
can describe quantum correlations by a classical model. The new Bell
inequality for three four-dimensional systems reads
\begin{eqnarray}
&&-5P(a_1+b_1+c_1=0)+P(a_1+b_1+c_1=1)+3P(a_1+b_1+c_1=2)+P(a_1+b_1+c_1=3)  \nonumber \\
&&+3P(a_1+b_1+c_2=0)-7P(a_1+b_1+c_2=1)+3P(a_1+b_1+c_2=2)+P(a_1+b_1+c_2=3) \nonumber \\
&&+3P(a_1+b_2+c_1=0)-7P(a_1+b_2+c_1=1)+3P(a_1+b_2+c_1=2)+P(a_1+b_2+c_1=3) \nonumber \\
&&+3P(a_2+b_1+c_1=0)-7P(a_2+b_1+c_1=1)+3P(a_2+b_1+c_1=2)+P(a_2+b_1+c_1=3) \nonumber \\
&&+3P(a_1+b_2+c_2=0)+P(a_1+b_2+c_2=1)-5P(a_1+b_2+c_2=2)+P(a_1+b_2+c_2=3) \nonumber \\
&&+3P(a_2+b_1+c_2=0)+P(a_2+b_1+c_2=1)-5P(a_2+b_1+c_2=2)+P(a_2+b_1+c_2=3) \nonumber \\
&&+3P(a_2+b_2+c_1=0)+P(a_2+b_2+c_1=1)-5P(a_2+b_2+c_1=2)+P(a_2+b_2+c_1=3) \nonumber \\
&&-P(a_2+b_2+c_2=0)-3P(a_2+b_2+c_2=1)-P(a_2+b_2+c_2=2)+5P(a_2+b_2+c_2=3)) \leq 12.
\label{bell1}
\end{eqnarray}

That the maximum value of the left hand side of inequality
(\ref{bell1}) for local theories is 12 can be given in the following
sense. By using $\sum_{r=0}^{3}P(a_i+b_j+c_k=r)=1$, the inequality
(\ref{bell1}) is reformed as
\begin{eqnarray}
&&-3P(a_1+b_1+c_1=0)+P(a_1+b_1+c_1=2)-5P(a_1+b_1+c_2=1)\nonumber \\
&&-P(a_1+b_1+c_2=3)-5P(a_1+b_2+c_1=1)-P(a_1+b_2+c_1=3)\nonumber \\
&&-5P(a_2+b_1+c_1=1)-P(a_2+b_1+c_1=3)+P(a_1+b_2+c_2=0)\nonumber \\
&&-3P(a_1+b_2+c_2=2)+P(a_2+b_1+c_2=0)-3P(a_2+b_1+c_2=2)\nonumber \\
&&+P(a_2+b_2+c_1=0)-3P(a_2+b_2+c_1=2)-P(a_2+b_2+c_2=1)\nonumber \\
&&+3P(a_2+b_2+c_2=3)) \leq 0.
\label{3qu4itinequality}
\end{eqnarray}
To beat the bound $0$, terms $P(a_1+b_1+c_1=2)$ and
$P(a_2+b_2+c_2=3)$ are taken equal to one first. This means that
$a_1+b_1+c_1+a_2+b_2+c_2=5$. Among the remained terms, we take
$P(a_1+b_1+c_2=3)$, $P(a_1+b_2+c_1=3)$ and $P(a_2+b_1+c_1=3)$ equal
to one to maximize the value of left hand side of the inequality
(\ref{3qu4itinequality}). As a result, $a_2+b_2+c_1=2$,
$a_2+b_1+c_2=2$ and $a_1+b_2+c_2=2$ according to the constraint
$a_1+b_1+c_1+a_2+b_2+c_2=5$. So
$P(a_1+b_1+c_2=1)=P(a_1+b_2+c_1=1)=P(a_2+b_1+c_1=1)=0$,
$P(a_2+b_2+c_1=2)=P(a_2+b_1+c_2=2)=P(a_1+b_2+c_2=2)=1$ and
$P(a_2+b_2+c_1=0)=P(a_2+b_1+c_2=0)=P(a_1+b_2+c_2=0)=0$. Therefore we
have $0 +1-0-1-0-1-0-1+0-3+0-3+0-3-0+3=-8\le 0$. If initially terms
$P(a_1+b_1+c_1=0)$ and $P(a_2+b_2+c_2=3)$ are taken equal to one
first. This means that $a_1+b_1+c_1+a_2+b_2+c_2=3$. Among the
remained terms, we take $P(a_1+b_1+c_2=3)$, $P(a_1+b_2+c_1=3)$ and
$P(a_2+b_1+c_1=3)$ equal to one to maximize the value of left hand
side of the inequality (\ref{3qu4itinequality}). As a result,
$a_2+b_2+c_1=0$, $a_2+b_1+c_2=0$ and $a_1+b_2+c_2=0$ according to
the constraint $a_1+b_1+c_1+a_2+b_2+c_2=3$. So
$P(a_1+b_1+c_2=1)=P(a_1+b_2+c_1=1)=P(a_2+b_1+c_1=1)=0$,
$P(a_2+b_2+c_1=2)=P(a_2+b_1+c_2=2)=P(a_1+b_2+c_2=2)=0$ and
$P(a_2+b_2+c_1=0)=P(a_2+b_1+c_2=0)=P(a_1+b_2+c_2=0)=1$. Therefore we
have $-3+0-0-1-0-1-0-1+1-0+1-0+1-0-0+3=0\leq 0$. Therefore, after
some simple and patient calculations, it can be shown that the
inequality (\ref{3qu4itinequality}) is always bounded by $0$ in a
local realistic model. Furthermore, the Bell inequality
(\ref{bell1}) is a tight inequality for three four-dimensional
systems \cite{Acin3}.

Let us now consider the maximum value that can be attained for the
inequality (\ref{bell1}) for quantum measurements on an entangled
quantum state. First, we specify the quantum state and measurement.
The initial state is a natural generalization of bipartite maximally
entangled state to three four-level systems,
\begin{eqnarray}
|\psi_4\rangle=\frac{1}{2}(|000\rangle+|111\rangle+|222\rangle+|333\rangle).
\end{eqnarray}
Consider a Gedanken experiment in which A, B and C measure
observables defined by unbiased symmetric multi-port beam splitters
\cite{Zukowski2} on $|\psi\rangle$. The unbiased symmetric
multi-port beam splitter is an optical device with $d$ input and $d$
output ports. In front of every input port there is a phase shifter
that changes the phase of the photon entering the given port. If a
phase shifter in some input port is set to zero and a photon enters
the device through this port then it has an equal chance of leaving
the device through any output port. The phase shifters can be
changed by the observers; they represent the local macroscopic
parameters available to the observers. The matrix elements of an
unbiased symmetric multi-port beam splitter are given by
$U_{kl}(\vec{\phi})=\frac{1}{\sqrt{d}}\alpha ^{kl}\exp(i\phi^{l})$,
where $\alpha=\exp(\frac{2i\pi}{d})$ and $\phi^{l}
(l=0,1,2,...,d-1)$ are the settings of the appropriate phase
shifters, for convenience we denote them as a $d$ dimensional vector
$\vec{\phi}=(\phi^0, \phi^1, \phi^2,..., \phi^{d-1})$. For
four-dimensional systems, $d=4$.

The quantum prediction for the probabilities of obtaining the
outcome $(a,b,c)$ is then given as
\begin{eqnarray}
P(a_i=a,b_j=b,c_k=c)=|\langle abc|U(\vec{\phi_A})\otimes U(\vec{\phi_B})\otimes U(\vec{\phi_C})|\psi_4\rangle|^2.
\end{eqnarray}
Thus the quantum analogue of the joint probability can be easily calculated
\begin{eqnarray}
&&P(a_i+b_j+c_k=r) \nonumber \\
&&=\frac{1}{16}\biggr(4+2\cos(\varphi^1-\varphi^0+\frac{\pi}{2}r)+2\cos(\varphi^2-\varphi^0+\pi r+2\cos(\varphi^2-\varphi^1+\frac{\pi}{2}r) \nonumber \\
&&+2\cos(\varphi^3-\varphi^0+\frac{3\pi}{2}r)
+2\cos(\varphi^3-\varphi^1+\pi
r)+2\cos(\varphi^3-\varphi^2+\frac{\pi}{2}r)\biggr),
\end{eqnarray}
where $\varphi^i=\phi^i_A+\phi^i_B+\phi^i_C,\; (i=0,1,2,3)$. In
order to look for the maximal violation of the inequality, we choose
the optimal settings as the following:
$\vec{\phi}_{A1}=\vec{\phi}_{B1}=\vec{\phi}_{C1}=(0,\frac{1}{3}\arccos(-\frac{1}{3}),\frac{1}{3}\arccos(-\frac{1}{3})-\frac{\pi}{3},\frac{\pi}{3})$,
$\vec{\phi}_{A2}=\vec{\phi}_{B2}=\vec{\phi}_{C2}=(0,\frac{1}{3}\arcsin\frac{7}{9},\frac{1}{3}\arcsin\frac{7}{9}+\frac{\pi}{6},-\frac{\pi}{6})$.
Numerical results show that for this choice, all the probability
terms have definite values as listed in Table \ref{tab1}.
 \begin{table}
\begin{tabular}{| c |c |c |c|}
\hline $p(a_1+b_1+c_1=0)$ & $p(a_1+b_1+c_1=1)$ & $p(a_1+b_1+c_1=2)$& $p(a_1+b_1+c_1=3)$  \\
\hline $0$ & $1/6$ & $2/3$ & $1/6$ \\
\hline $p(a_1+b_1+c_2=0)$ & $p(a_1+b_1+c_2=1)$ & $p(a_1+b_1+c_2=2)$& $p(a_1+b_1+c_2=3)$  \\
\hline $1/2$ & $0$ & $1/2$ & $0$\\
\hline $p(a_1+b_2+c_1=0)$ & $p(a_1+b_2+c_1=1)$ & $p(a_1+b_2+c_1=2)$& $p(a_1+b_2+c_1=3)$  \\
\hline $1/2$ & $0$ & $1/2$ & $0$\\
\hline $p(a_2+b_1+c_1=0)$ & $p(a_2+b_1+c_1=1)$ & $p(a_2+b_1+c_1=2)$& $p(a_2+b_1+c_1=3)$  \\
\hline $1/2$ & $0$ & $1/2$ & $0$\\
\hline $p(a_1+b_2+c_2=0)$ & $p(a_1+b_2+c_2=1)$ & $p(a_1+b_2+c_2=2)$& $p(a_1+b_2+c_2=3)$  \\
\hline $2/3$ & $1/6$ & $0$ & $1/6$ \\
\hline $p(a_2+b_1+c_2=0)$ & $p(a_2+b_1+c_2=1)$ & $p(a_2+b_1+c_2=2)$& $p(a_2+b_1+c_2=3)$  \\
\hline $2/3$ & $1/6$ & $0$ & $1/6$ \\
\hline $p(a_2+b_2+c_1=0)$ & $p(a_2+b_2+c_1=1)$ & $p(a_2+b_2+c_1=2)$& $p(a_2+b_2+c_1=3)$  \\
\hline $2/3$ & $1/6$ & $0$ & $1/6$ \\
\hline $p(a_2+b_2+c_2=0)$ & $p(a_2+b_2+c_2=1)$ & $p(a_2+b_2+c_2=2)$& $p(a_2+b_2+c_2=3)$  \\
\hline $1/18$ & $0$ & $1/18$ & $8/9$ \\
\hline
\end{tabular}
 \caption{The values of the quantum joint probabilities in inequality (\ref{bell1}) with appropriate angle settings.}\label{tab1}
\end{table}
Putting them into the left hand side of the inequality
(\ref{bell1}), we arrive at
$2\frac{1}{6}+3\frac{2}{3}+3(3\frac{1}{2}+3\frac{1}{2})
+3(3\frac{2}{3}+2\frac{1}{6})-2\frac{1}{18}+5\frac{8}{9}
=\frac{68}{3}>12$.

In Ref. \cite{v}, a proposal was made to measure the strength of
violation of local realism by the minimal amount of noise that must
be added to the system in order to hide the non-classical character
of the observed correlations. This is equivalent to a replacement of
the pure state $|\psi\rangle\langle\psi|$ by the mixed state
$\rho(F)$ of the form
$\rho(F)=(1-F)|\psi\rangle\langle\psi|+\frac{F}{56}I\otimes I\otimes
I$, where $I$ is an identity matrix and $F (0\leq F\leq 1)$ is the
amount of noise present in the system. For $F=0$, local realistic
description does not exist, whereas it does for $F=1$. Therefore,
there exists some threshold value of $F$, denoted by $F_{thr}$, such
that for every $F\leq F_{thr}$, local and realistic description does
not exist. The threshold fidelity for the three four-level systems
is determined by $(1-F_{thr})\frac{68}{3}=12$, namely
$F_{thr}=\frac{8}{17}=0.4706$.

Similarly, we propose a Bell inequality for three five-dimensional
systems based on the Gedanken experiment:
\begin{eqnarray}
&&-2P(a_1+b_1+c_1=0)+P(a_1+b_1+c_1=1)+P(a_1+b_1+c_1=4)  \nonumber \\
&&+P(a_1+b_1+c_2=0)-2P(a_1+b_1+c_2=2)+P(a_1+b_1+c_2=4) \nonumber \\
&&+P(a_1+b_2+c_1=0)-2P(a_1+b_2+c_1=2)+P(a_1+b_2+c_1=4) \nonumber \\
&&+P(a_2+b_1+c_1=0)-2P(a_2+b_1+c_1=2)+P(a_2+b_1+c_1=4) \nonumber \\
&&+P(a_1+b_2+c_2=0)+P(a_1+b_2+c_2=3)-2P(a_1+b_2+c_2=4) \nonumber \\
&&+P(a_2+b_1+c_2=0)+P(a_2+b_1+c_2=3)-2P(a_2+b_1+c_2=4) \nonumber \\
&&+P(a_2+b_2+c_1=0)+P(a_2+b_2+c_1=3)-2P(a_2+b_2+c_1=4) \nonumber \\
&&-2P(a_2+b_2+c_2=1)+P(a_2+b_2+c_2=3)+P(a_2+b_2+c_2=4) \leq 4,
\label{5bell}
\end{eqnarray}
which is satisfied by the local and realistic theories. By the way,
the Bell inequality (\ref{5bell}) is also tight \cite{Acin3}.

Using specified quantum state and measurement, we calculate the
maximum value that can be attained for the inequality (\ref{5bell}).
The considered state is  a natural generalization of bipartite
maximally entangled state to three five-level systems,
\begin{eqnarray}
|\psi_5\rangle=\frac{1}{\sqrt{5}}(|000\rangle+|111\rangle+|222\rangle+|333\rangle+|444\rangle).
\end{eqnarray}
The measurement is also based on unbiased symmetric multi-port beam
splitter with $d=5$ and the quantum prediction for the probabilities
of obtaining the outcome $(a,b,c)$ is then given as
\begin{eqnarray}
P(a_i=a,b_j=b,c_k=c)=|\langle abc|U(\vec{\phi_A})\otimes
U(\vec{\phi_B})\otimes U(\vec{\phi_C})|\psi_5\rangle|^2.
\end{eqnarray}
Thus the quantum analogue of the joint probability is given as
\begin{eqnarray}
&&P(a_i+b_j+c_k=r) \nonumber \\
&&=\frac{1}{25}\biggr(5+2\cos(\varphi^1-\varphi^0+\frac{2\pi}{5}r)+2\cos(\varphi^2-\varphi^0+\frac{4\pi}{5}r)+2\cos(\varphi^2-\varphi^1+\frac{2\pi}{5}r) \nonumber \\
&&+2\cos(\varphi^3-\varphi^0+\frac{6\pi}{5}r)+2\cos(\varphi^3-\varphi^1+\frac{4\pi}{5}r)+2\cos(\varphi^3-\varphi^2+\frac{2\pi}{5}r) \nonumber \\
&&+2\cos(\varphi^4-\varphi^0+\frac{8\pi}{5}r)+2\cos(\varphi^4-\varphi^1+\frac{6\pi}{5}r)
+2\cos(\varphi^4-\varphi^2+\frac{4\pi}{5}r)+2\cos(\varphi^4-\varphi^3+\frac{2\pi}{5}r)\biggr),
\end{eqnarray}
where $\varphi^i=\phi^i_A+\phi^i_B+\phi^i_C \; (i=0,1,2,3,4)$.
Numerical results show that for the following angle settings
$\vec{\phi}_{A1}=\vec{\phi}_{B1}=\vec{\phi}_{C1}=(0,\beta_1,\beta_2,-\beta_2,-\beta_1)$,
$\vec{\phi}_{A2}=\vec{\phi}_{B2}=\vec{\phi}_{C2}=(0,\beta_1+\frac{\pi}{5},\beta_2+\frac{2\pi}{5},-\beta_2-\frac{2\pi}{5},-\beta_1-\frac{\pi}{5})$,
where $\cos(3\beta_1)-\cos(3\beta_2)=\frac{1}{2}$, the maximum value
of the left hand side of inequality (\ref{5bell}) attained is
$6.72216$. The threshold fidelity for the three five-level systems
is determined by $(1-F_{thr})\times 6.72216 =4$, namely
$F_{thr}=0.40495$.

Given the above known Bell inequalities for three particles with
$d=2, 3, 4, 5$, it is worthy of noting that such tight inequalities
exhibit perfect symmetries. One symmetry is that these Bell
inequalities are symmetric under the permutations of subsystems $A,
B,$ and $C$. The second symmetry is that they can be expressed in a
general form. Based on which, the structure of tight Bell
inequalities is suggested as
\begin{eqnarray}
&&\frac{1}{d(d-1)}\sum_{ijk}\sum_{r=0}^{d-1} f_{d;ijk}^r P(a_i+b_j+c_k=r) \leq 1,
\end{eqnarray}
with $-1 \leq \frac{f_{d;ijk}^r}{d(d-1)} \leq 1$. It is worthy of
noting that the tight Bell inequalities for two qudits (i.e., the
CGLMP inequality, see \cite{Gisin}) can be written in a similar
form,
\begin{eqnarray}
&&\frac{1}{d-1}\sum_{ij}\sum_{r=0}^{d-1} f_{d;ij}^r P(a_i+b_j=r) \leq 1,
\end{eqnarray}
with $-1 \leq \frac{f_{d;ij}^r}{d-1} \leq 1$. The coefficients
$f_{d;ijk}^r$ and $f_{d;ij}^r$ are integers or half integers. Note
that these inequalities are fulfilled with displacement of
probabilities. In other words, for a known number $m$, where $m \leq
d-1$, the above inequalities are still true for local realistic
description by replacing $P(a_i+b_j+c_k=r)$ with
$P(a_i+b_j+c_k=r+m)$, which is the third symmetry of the set of Bell
inequalities for three particles.

\section{ Interesting Bell inequalities of three-qubit reduced from those of
three-qudit}

 In 1991 Gisin presented a theorem, which states that {\it
any} pure entangled state of two particles violates a Bell
inequality for two-particle correlation functions
\cite{gisin,popescu}. Recent investigations show a surprising result
that there exists a family of pure entangled $N$-qubit states that
does not violate any Bell inequality for $N$-particle correlations
for the case of a standard Bell experiment on $N$ qubits
\cite{scarani}. This family is the generalized GHZ states given by
\begin{eqnarray}
|\psi\rangle_{GHZ}=\cos\xi|0\cdots 0\rangle+\sin\xi|1\cdots
1\rangle,
 \label{eq1}
\end{eqnarray}
with $0 \le \xi \le \pi/4$. The usual GHZ states \cite{ghz} are for
$\xi=\pi/4$. For a three-qubit system, whose corresponding
generalized GHZ state reads
$|\psi\rangle_{GHZ}=\cos\xi|000\rangle+\sin\xi|111\rangle$, it has
been shown that for the region $\xi \in (0, \pi/12]$, the
inequalities given in \cite{ZB} are not violated based on the
standard Bell experiment. Recently, we developed a three-qubit Bell
inequality which is a solution to such a problem. That is all pure
entangled states of three qubits violates the Bell inequality given
in \cite{3qubit}. Indeed Bell inequalities are sensitive to the
presence of noise and above a certain amount of noise the Bell
inequalities will cease to be violated by a quantum system. However,
it seems that the inequality in Ref. \cite{3qubit} is not good
enough to the resistance of noise. For the three-qubit GHZ state,
the threshold visibility is $V_{GHZ}=4\sqrt{3}/9=0.7698$ and for the
W state, the threshold visibility is $V_{W}=0.7312$. Tne inequality
in Ref. \cite{3qubit} can be derived from three-qutrit Bell
inequality (\ref{3qutritinequality}). Actually any Bell inequality
for tripartite $d\;(d>2)$-level systems may reduce to a Bell
inequality for three qubits when one considers only two outcomes of
measurement.

Here we present a new Bell inequality for three-qubit systems which
is reduced from inequality (\ref{bell1})
\begin{eqnarray}
&&3P(a_1+b_1+c_1=0)+P(a_1+b_1+c_1=1)-5P(a_1+b_1+c_1=2)+P(a_1+b_1+c_1=3)  \nonumber \\
&&+3P(a_1+b_1+c_2=0)+P(a_1+b_1+c_2=1)+3P(a_1+b_1+c_2=2)-7P(a_1+b_1+c_2=3) \nonumber \\
&&+3P(a_1+b_2+c_1=0)+P(a_1+b_2+c_1=1)+3P(a_1+b_2+c_1=2)-7P(a_1+b_2+c_1=3) \nonumber \\
&&+3P(a_2+b_1+c_1=0)+P(a_2+b_1+c_1=1)+3P(a_2+b_1+c_1=2)-7P(a_2+b_1+c_1=3) \nonumber \\
&&-5P(a_1+b_2+c_2=0)+P(a_1+b_2+c_2=1)+3P(a_1+b_2+c_2=2)+P(a_1+b_2+c_2=3) \nonumber \\
&&-5P(a_2+b_1+c_2=0)+P(a_2+b_1+c_2=1)+3P(a_2+b_1+c_2=2)+P(a_2+b_1+c_2=3) \nonumber \\
&&-5P(a_2+b_2+c_1=0)+P(a_2+b_2+c_1=1)+3P(a_2+b_2+c_1=2)+P(a_2+b_2+c_1=3) \nonumber \\
&&-P(a_2+b_2+c_2=0)+5P(a_2+b_2+c_2=1)-P(a_2+b_2+c_2=2)-3P(a_2+b_2+c_2=3)) \leq 12.
\label{bell2}
\end{eqnarray}
The inequality can be expressed in terms of correlation functions
\begin{eqnarray}
&&-E(A_1B_1C_1)+E(A_1B_1C_2)+E(A_1B_2C_1)+E(A_2B_1C_1)-E(A_2B_2C_2) \nonumber \\
&&-E(A_1B_2)-E(A_2B_1)-E(A_2B_2)-E(A_1C_2)-E(A_2C_1)-E(A_2C_2)\nonumber \\
&&-E(B_1C_2)-E(B_2C_1)-E(B_2C_2)+E(A_1)+E(B_1)+E(C_1)\leq 3.
\label{second}
\end{eqnarray}

\begin{figure}
   \includegraphics[width=8cm]{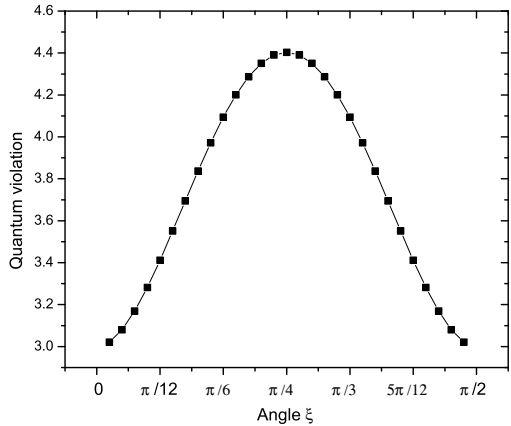}\\
\caption{Numerical results for the generalized GHZ states
$|\psi_2\rangle_{\rm GHZ}=\cos\xi|000\rangle+\sin\xi|111\rangle$,
which violate the inequality (\ref{second}) except
$\xi=0,\pi/2$.}\label{3qubit2GHZ}
\end{figure}

\begin{figure}
   \includegraphics[width=8cm]{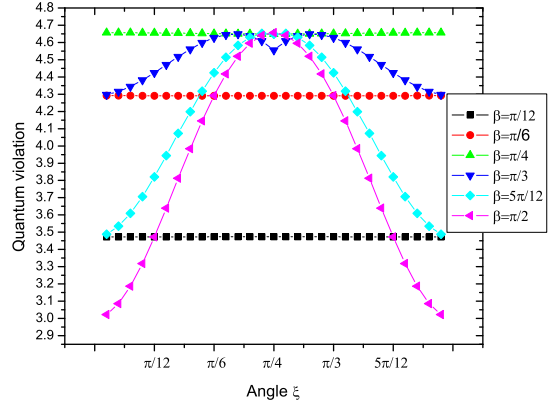}\\
\caption{Numerical results for the generalized W states
$|\psi\rangle_{W}=\sin\beta\cos\xi|100\rangle+\sin\beta\sin\xi|010\rangle+\cos\beta|001\rangle$
which violate the inequality (\ref{second}) for different $\xi$ and
$\beta$. Here the cases $\beta=\pi/12, \pi/6,\pi/4,\pi/3,5\pi/12$,
and $\pi/2$ are considered.} \label{3qubit2W}
\end{figure}

The above inequality (\ref{second}) includes the terms of single
correlation functions, it is symmetric under the permutations of
$A_j, B_j$ and $C_j$. Quantum mechanically, the above inequality is
violated by all pure entangled states of three qubits. Pure states
of three qubits constitute a five-parameter family, with equivalence
up to local unitary transformations. This family has the following
representation \cite{Acin00}
\begin{eqnarray}
|\psi\rangle & = & \sqrt{\mu_0}|000\rangle+\sqrt{\mu_1}e^{i\phi}|100\rangle
+\sqrt{\mu_2}|101\rangle \nonumber\\
&& +\sqrt{\mu_3}|110\rangle+\sqrt{\mu_4}|111\rangle,
\end{eqnarray}
with $\mu_i \ge 0$, $\sum_i \mu_i=1$, and $0 \le \phi \le \pi$.
Numerical results show that this Bell inequality for probabilities
is violated by all pure entangled states of three-qubit systems.
However, no analytical proof of the conclusion has been given. In
the following, some special cases will be given to show the
inequality (\ref{second}) is violated by all pure entangled states.
The first quantum state considered is generalized GHZ state
$|\psi_2\rangle_{\rm GHZ}=\cos\xi|000\rangle+\sin\xi|111\rangle$.
The inequality (\ref{second}) is violated by the generalized GHZ
states for the whole region except $\xi=0,\pi/2$. For the GHZ state
with $\xi=\pi/4$, the quantum violation reaches its maximum value
$4.40367$. The variation of the violation with $\xi$ is shown in
Fig. \ref{3qubit2GHZ}. Another state considered is generalized W
state
$|\psi\rangle_{W}=\sin\beta\cos\xi|100\rangle+\sin\beta\sin\xi|010\rangle+\cos\beta|001\rangle$.
By fixing the value of $\beta$, quantum violation of the inequality
(\ref{second}) varies with $\xi$ (see Fig. \ref{3qubit2W}). The
inequality (\ref{second}) is violated by the generalized W states
except the cases with $\beta=\frac{\pi}{2}$, $\xi=0$ and
$\xi=\frac{\pi}{2}$. The states in these cases are direct-product
states which do not violated any Bell inequality. For the standard W
state, quantum violation of the inequality (\ref{second}) approaches
$4.54086$.

\begin{figure}
   \includegraphics[width=8cm]{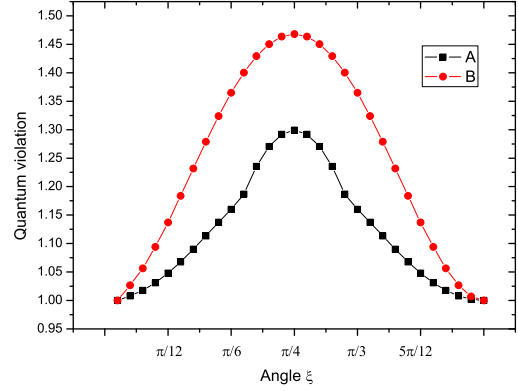}\\
\caption{Violation of two Bell inequalities for the generalized GHZ
state of three qubits with different value of $\xi$, where curve A
is for inequality given in Ref. \cite{3qubit} and curve B is for a
new inequality (\ref{second}).} \label{fig}
\end{figure}

Hence the inequality (\ref{second}) is also one candidate to
generalize the theorem of Gisin to three-qubit systems. One of the
features of the new inequality for three qubits is that it is highly
resistant to noise. The inequality (\ref{second}) is violated by the
generalized GHZ state
$|\psi_2\rangle_{GHZ}=\cos\xi|000\rangle+\sin\xi|111\rangle$ for the
whole region, the threshold visibility is $V_{thr}^{GHZ} = 0.68125$.
The inequality (\ref{second}) is also violated by the W state, the
threshold visibility is $V_{thr}^{W} =0.660668$. We plot the
variation of quantum violation for the generalized GHZ states with
angle $\xi$ for inequality (\ref{second}) and inequality given in
Ref. \cite{3qubit}, see Fig. \ref{fig}. In plotting the figure, we
reform the expressions of these two inequalities as
\begin{eqnarray}
\frac{1}{4}&[&Q(A_1B_1C_1)-Q(A_1B_2C_2)-Q(A_2B_1C_2)-Q(A_2B_2C_1)+2Q(A_2B_2C_2) \nonumber \\
&&-Q(A_1B_1)-Q(A_1B_2)-Q(A_2B_1)-Q(A_2B_2)+Q(A_1C_1)+Q(A_1C_2) \nonumber \\
&&+Q(A_2C_1)+Q(A_2C_2)+Q(B_1C_1)+Q(B_1C_2)+Q(B_2C_1)+Q(B_2C_2)] \leq
1,
\end{eqnarray}
\begin{eqnarray}
\frac{1}{3}&[&-Q(A_1B_1C_1)+Q(A_1B_1C_2)+Q(A_1B_2C_1)+Q(A_2B_1C_1)-Q(A_2B_2C_2) \nonumber \\
&&-Q(A_1B_2)-Q(A_2B_1)-Q(A_2B_2)-Q(A_1C_2)-Q(A_2C_1)-Q(A_2C_2)\nonumber \\
&&-Q(B_1C_2)-Q(B_2C_1)-Q(B_2C_2)+Q(A_1)+Q(B_1)+Q(C_1)]\leq 1,
\end{eqnarray}
respectively. By the reformation, the violation degrees of the two
inequalities can be compared directly. Comparing the results of the
inequality given in Ref. \cite{3qubit}, the new inequality
(\ref{second}) is indeed more resistant to noise. It seems that we
could derive some new three-qubit Bell inequalities, which would be
more highly resistance to noise, if we get other Bell inequalities
of tripartite $d$-dimensional ($d>4$) quantum systems.

When setting $C_1=-1, C_2=1$, the inequality (\ref{second}) reduces
directly to the CHSH inequality for two-qubit
      $$ E(A_1B_1)-E(A_1B_2)-E(A_2B_1)-E(A_2B_2) \leq 2,$$
which is equivalent to $$ E(A_1B_1)+E(A_1B_2)+E(A_2B_1)-E(A_2B_2) \leq 2. $$

Starting from the Bell inequality for three five-level systems,
another three-qubit Bell inequality can be obtained as
\begin{eqnarray}
&&P(a_1+b_1+c_1=1)-2P(a_1+b_1+c_1=2)+P(a_1+b_1+c_1=3)+P(a_1+b_1+c_2=1)+P(a_1+b_1+c_2=2)  \nonumber \\
&&+P(a_1+b_2+c_1=1)+P(a_1+b_2+c_1=2)+P(a_2+b_1+c_1=1)+P(a_2+b_1+c_1=2) \nonumber \\
&&+P(a_1+b_2+c_2=0)-2P(a_1+b_2+c_2=1)+P(a_1+b_2+c_2=2)+P(a_2+b_1+c_2=0) \nonumber \\
&&-2P(a_2+b_1+c_2=1)+P(a_2+b_1+c_2=2)+P(a_2+b_2+c_1=0)-2P(a_2+b_2+c_1=1) \nonumber \\
&&+P(a_2+b_2+c_1=2)+P(a_2+b_2+c_2=0)+P(a_2+b_2+c_2=1)-2P(a_2+b_2+c_2=3) \leq 4.
\label{redu5b}
\end{eqnarray}
The above inequality is just the Mermin inequality when it is
expressed in terms of correlation functions:
\begin{eqnarray}
-Q_{111}+Q_{122}+Q_{212}+Q_{221}\leq 2.
\end{eqnarray}

\section{Summary}

To summarize, we have presented the tight Bell inequalities
expressed by probabilities for three four- and five-dimensional
systems. The tight structure of Bell inequalities for three
$d$-dimensional systems (qudits) is also proposed. Moreover, a new
Bell inequality for three qubits is derived (or say, reduced) from
the inequality for three four-level systems. The inequality
(\ref{second}) is violated by any pure three-qubit entangled state,
and it is more resistant to noise compared with the one given in
Ref. \cite{3qubit}. Furthermore, the tight Mermin inequality for
thre-qubit can be obtained by reducing the Bell inequality for three
five-level systems [see Eq. (\ref{5bell}) and Eq. (\ref{redu5b})].

\vspace{2mm}

 {\bf ACKNOWLEDGMENTS} The authors thank A. Acin for many
useful discussions. This work was supported by NUS research Grant
No. WBS: R-144-000-123-112. J.L.C acknowledges financial supports
from NSF of China (Grant No. 10605013), Program for New Century
Excellent Talents in University, and the Project-sponsored by SRF
for ROCS, SEM. C.F.W. acknowledges financial support from Singapore
Millennium Foundation.


\begin{thebibliography}{99}
\bibitem{Acin3} Private communication with A. Acin, who has performed
the culculation and proved that the Bell inequalities (\ref{bell1})
and (\ref{5bell}) were tight.
\end{thebibliography}
\end{document}